# Understanding Substrate Effects on Two-Dimensional MoS$_2$ Growth: a Kinetic Monte Carlo Approach

*Samuel Aldana\*, Ion Spiridon, Lulin Wang and Hongzhou Zhang\**

S. Aldana, I. Spiridon, L. Wang, H. Zhang

Centre for Research on Adaptive Nanostructures and Nanodevices (CRANN) and Advanced Materials and Bioengineering Research (AMBER) Research Centers, Trinity College Dublin, Dublin 2, Ireland.

School of Physics, Trinity College Dublin, Dublin 2, Ireland

E-mail: aldanads@tcd.ie; hozhang@tcd.ie

ABSTRACT: Controlling the morphology of two-dimensional (2D) transition metal dichalcogenides (TMDs) plays a key role in their applications. Although chemical vapor deposition can achieve wafer-scale growth of 2D TMDs, a comprehensive theoretical framework for effective growth optimization is lacking. Atomistic modeling methods offer a promising approach to delve into the intricate dynamics underlying the growth. In this study, we employ kinetic Monte Carlo (kMC) simulations to identify crucial parameters that govern the morphology of MoS$_2$ flakes grown on diverse substrates. Our simulations reveal that large adsorption rates significantly enhance growth speed, which however necessitates rapid edge migration to achieve compact triangles. Substrate etching can tune the adsorption-desorption process of adatoms and enable preferential growth within a specific substrate region, controlling the flake morphology. This study unravels the complex dynamics governing 2D TMD morphology, offering a theoretical framework for decision-making in the design and optimization of TMD synthesis processes.

# 1. Introduction

Single or few-layer transition metal dichalcogenides (TMDs), exemplified by $MoS_2$, have garnered significant research attention due to their exceptional properties.[1-4] These materials stand as promising contenders for the next generation of electronic devices.[5-11] High-quality thin films are not only needed for the fundamental understanding of new materials but also demanded for industry applications.[12] The mechanical exfoliation is a well-established technique for rapid prototyping and proof-of-concept experiments.[3, 7, 13-15] However, it proves impractical for large-scale production. In contrast, "bottom-up" techniques like molecular beam epitaxy (MBE)[16-17] and chemical vapor deposition (CVD)[18-19] can be integrated into large-scale semiconductor processes. Nevertheless, to ensure the reliability of device fabrication, significant advancements are required in these "bottom-up" approaches to achieve desirable film morphologies. For example, it is imperative to control their microstructures such as thickness, grain size, grain boundaries, and defect density to enhance device reliability and reproducibility. [20-23] Wafer-scale growth of single crystal 2D TMD films has been accomplished on special substrates,[18, 24-26] and the control of the film morphology has been widely attempted.[12, 18, 27] Anisotropic growth, induced both by the substrate[28] and the use of alkali[29], has been documented. It has been elucidated that the flakes exhibit branched and defective morphology when the edge diffusion rate is relatively low compared with the attachment rate of adatoms to the flake.[12, 27, 30] The variation in the precursor ratio (chalcogen/metal) also substantially affects flake morphology, leading to different shapes such as triangular, hexagonal or even anisotropic growth.[31-34] Additionally, higher adatom density and diffusion rates are conducive to more frequent nucleation events, impacting the average size of the flakes.[12] Nevertheless, there still exist numerous challenges in modulating the flake morphology. CVD 2D films often manifest high defect densities with reduced crystallinity in contrast to their exfoliated counterparts. Particularly, branched flakes with discontinuous structures, which are encountered quite often in CVD growth, can introduce inconsistencies and unpredictability in material properties[27, 31, 35]

A prominent challenge in optimizing experimental parameters for CVD growth lies in the absence of a theoretical framework that can elucidate the fundamental mechanisms which govern 2D domain morphology under various growth conditions. This challenge is compounded by the multitude of kinetic processes involved and the wide spectrum of growth conditions encompassing atomic flux, temperature, and substrate conditions. These complexities hinder the identification of pivotal factors governing the morphology of the film. Furthermore, the laborious and time-consuming nature of experimental investigation of the growth mechanisms, coupled with the difficulty of establishing direct correlations between experimental parameters and atomic-scale kinetic processes, exacerbates this issue. Theoretical simulations can gain insights into the growth process, facilitate the analysis of experimental findings, and identify the critical variables relevant to growth optimization. Several methods have been employed to understand the growth of TMDs and multiscale models have been proposed[36-38]. First-principle techniques exemplified by the density functional theory (DFT)[39-42] are effective for scrutinizing processes involving small volumes at short time scales such as surface diffusion, edge diffusion, and atomic adsorption/desorption,[27] but DFT-based methods are not suited for replicating dynamic processes significantly deviating from equilibrium on a macroscopic scale. Molecular dynamics also proves to be a potent technique for investigating the impact of the substrate on $MoS_2$ morphology[43], yet encounters limitations in reaching macroscopic time scales. Furthermore, it should be noted that modeling the crystal growth of 2D TMDs necessitates an open system where matter can be introduced and removed. Alternative approaches, such as the phase field method,[37, 44-45] are apt to model the 2D material growth over large time and length scales, capturing complex morphologies in the mesoscale. However, these methods offer a continuum description of the evolving material, overlooking stochastic phenomena and microscopic details. The kMC method is a suitable approach to studying processes at a macroscopic time and spatial scale inherent to crystal growth since kMC enables stochastic and atomistic simulations that bridge the microscopic evolution of the system and the macroscopic outcomes, elucidating mechanisms often challenging to observe experimentally[46-48]. Notably, the kMC method has been a valuable tool to reveal various aspects of the CVD, offering insights into phenomena such as the formation of compact triangular islands,[49] the transition to fractal islands and the intricate competition among the kinetic processes of the growth of diverse 2D materials, e.g. graphene,[50] few-layer $MoS_2$,[30] and monolayer $WSe_2$[27]. Additionally, the method has also unraveled the effects of structural defects in 2D TMDs [51-52].

However, it is noteworthy that these previous studies have focused on flake morphology grown on a uniform substrate. In contrast, we employ kMC simulations to investigate the dynamics of the formation of $MoS_2$ monolayers at the interface between different substrates in the CVD growth. Our investigation delves into the intricate interplay of several crucial factors and their influence on flake morphology in a two-domain scenario, including adsorption rates, and activation energies associated with adatom desorption, on-substrate adatom migration, and edge migration. As such, we analyzed the morphological transformation occurring in 2D $MoS_2$ flakes induced by substrate roughness. We corroborate the robustness of our findings with experimental observations.

## 2. Simulation description

The simulation is designed to replicate the growth of triangular $MoS_2$ flakes on a $Si/SiO_2$ substrate. The substrate has been patterned by plasma etching, which alters the roughness of designated regions of the substrate. Details about the etching process and the CVD growth can be found in Supporting Information, Section 1. We employ a rejection-free kinetic Monte Carlo physical simulator, a methodology comprising two fundamental steps: (1) the computation of the transition rates associated with potential events, and (2) the selection of one of these events using a single random number.[46] The model is designed to simulate the growth of $MoS_2$ on a simulation domain of $50 \times 50$ nm and a lattice constant $a = 320$ pm (see **Figure 1a**). Our simulations commence with a triangular seed with zigzag S-edge (see the black triangle in Figure 1a), as it is one of the most thermodynamically stable configurations.[41] Figure 1b shows a SEM image of a typical CVD growth of $MoS_2$ with triangular-shaped flakes. The introduction of additional flakes is excluded. Figure 1c illustrates the key growth processes relevant to the simulation, which involve events characterized by distinct activation energies. Specifically, it includes the adsorption of S adatoms onto the substrate (i.e., event 1 in Figure 1c) and the desorption of the adatoms back into the gas phase (event 2). The adsorption rate, $r_a$, measured in monolayers per milisecond (ML ms$^{-1}$), represents the rate at which atoms are deposited onto the substrate. It signifies the rate of adatoms introduced into the simulation domain. These adatoms can diffuse to any of their six nearest unoccupied neighboring positions or desorb with an energy barrier of 0.6 eV. Figure 1d shows the key diffusion processes: Event 3 represents the diffusion of adatoms on the substrate with an energy barrier of 0.4 eV and Event 5 the migration of adatoms along the growth front with an activation energy of 1 eV[27]. The adatoms can bond with the crystal, i.e., Event 4. Notably, the activation energy for the S adatom to be bonded, particularly in the zigzag direction, is 0.44 eV, while growth in the armchair direction requires an activation energy of 0.7 eV.[53-54] In our simulation, we assume the Mo source is sufficient, ensuring that once an S adatom is attached to the crystal edge, the closest Mo sites are promptly filled.[30] To investigate the influence of varying precursor proportions on flake morphology, the simulations should explicitly incorporate the processes involving Mo (e.g. adatoms adsorption-desorption and migration) and the interaction with S. The simulation is deemed complete when the 2D flake covers 20% of the simulation domain.

The rates of aforementioned events are determined by the transition state theory dependent on the temperature and the activation energy, i.e., $\Gamma = \nu \cdot \exp(-E_A/K_B T)$, where $\nu = 7 \times 10^{13}$ s$^{-1}$ is the attempt frequency, $K_B$ is the Boltzmann constant, $T = 1108$ K the growth temperature used in the experiment and $E_A$ the activation energy of a given event. It is important to note that the temperature homogenizes the contributions of the different processes by promoting their probability. The transition rates are arranged in a balanced binary tree, enabling the utilization of a binary search method to efficiently locate the randomly chosen event through the kMC technique.[46-48] Within this data structure, each node at the lowest level corresponds to an individual event, while nodes at higher levels represent the cumulative sum of values from their respective child nodes. At the topmost node, the sum of all processes is stored. By employing this sorted data structure, the search process effectively eliminates either the left or right branch at each step, resulting in improved computational efficiency. The time step is calculated with a second random number as follows: $t = -\ln(rand)/\sum \Gamma$, where $rand$ stands for a random number between 0 and 1 and $\sum \Gamma$ is the summation of the possible events.

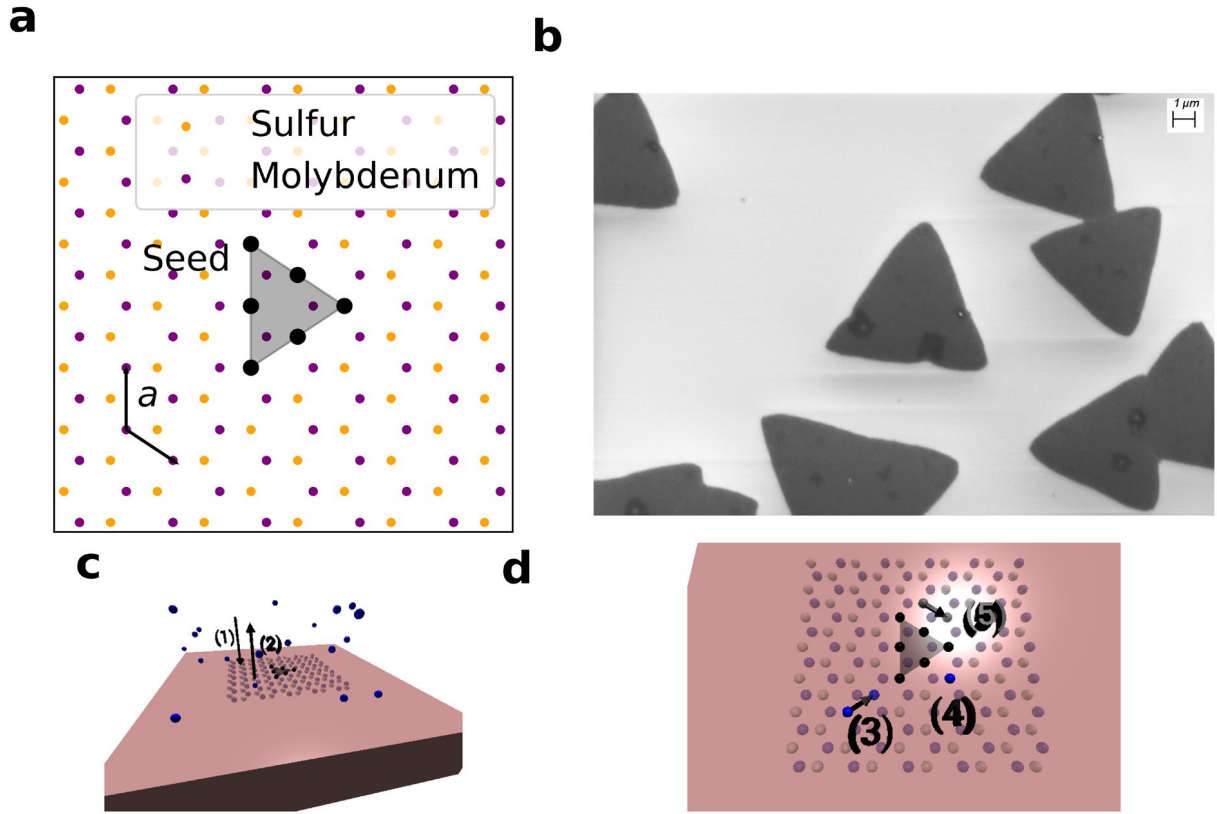

**Figure 1: Simulation domain, processes and experimental flakes.** a) Schematic representation of the 2H-MoS$_2$ crystal lattice, exhibiting a hexagonal structure with a lattice constant $a = 320$ pm. Purple particles represent Mo positions, while orange particles denote S positions. The black-shaded region indicates the crystal seed, featuring a zigzag-S edge, with black particles representing S belonging to the crystal. b) Typical SEM image of a CVD growth of MoS$_2$ resulting in triangular-shaped flakes. c) Schematic illustration of the adsorption (1) – desorption (2) equilibrium between gas-phase atoms and absorbed adatoms on the substrate, indicated by blue particles. The color scheme corresponds to that used in a), with the crystal seed represented by black particles and the substrate in orange. The semi-transparent particles symbolize potential growth units of the crystal. d) Schematic representation of adatom migration on the substrate (ad3), potentially leading to the crystal's edge and, subsequently, to incorporation into the flake domain (4). Atoms within the flake may also migrate along the edge (5) until they reach a stable position by clustering with other atoms.

## 3. Results and discussion

Our simulation first replicates the morphology control demonstrated in the experiment where the substrate surface is modified by the plasma etching. The modification can significantly alter locally the key processes of the 2D growth, i.e., the adsorption and desorption rate as well as the activation energies governing atom migration. These growth processes can be experimentally controlled by the adjustments in precursor concentrations, modulation of flow rates, as well as the choice of a substrate and application of substrate treatments (e.g., the named plasma etching or a functionalization). In this work, we simulate the impact the plasma etching might have varying the transition rates of these processes. Figure 2a illustrates that the growth speed ($v$) increases as a $v \propto \exp(1.09 \times r_a)$. The typical morphologies are depicted in the right panel of Figure 2a, corresponding to the points (I-VI). Flakes at points I and II maintain a compact triangular shape (denoted by cyan symbols in Figure 2), while with higher adsorption rates (as observed in III-IV) and hence larger growth speed ($v$), their shapes progressively deteriorate and become irregular (denoted by orange symbols) or branched (denoted by blue symbols). When the desorption rate remains unchanged, a higher adsorption rate ($r_a$) results in a higher adatom density, which enhances the joining rate of adatoms to the crystal and the growth speed ($v$). A low adatom flux allows atoms at the edge of the flake to migrate until they reach a stable position, and this reshaping process results in the formation of a compact triangle. Conversely, when the joining rate is high, atoms at the edge join and develop into branches suppressing the reshaping process.[12, 27] Figure 2b shows the effect of edge migration energy on the flake morphology where a larger symbol signifies a larger migration energy. Lowering the activation energy expedites the reshaping. This indicates a growth strategy to achieve large and triangular flakes, i.e., large compact triangles may be achievable under high surface adsorption rates with enhanced edge migration. In the inset of Figure 2b, we illustrate the adsorption rate limit ($r_a^{lim}$) necessary to achieve a compact triangle for various activation energies of edge migration on the left axis, while the corresponding growth speed on the right axis. For example, we can significantly increase the growth speed ($v$) by increasing the adsorption rate ($r_a$) by more than two orders of magnitude, all while retaining the capability to produce compact triangles. This is achievable by reducing the activation energy for edge migration from 1.2 to 0.6 eV. In Figure 2c, we explore the effects of modifying the adsorption-desorption equilibrium by altering the activation energy for adatom desorption. Lower activation energies result in a decreased adatom density on the substrate, diminishing the likelihood of adatoms migrating to the crystal edge before returning to the gas phase. Conversely, higher activation energies extend the duration during which adatoms remain bound to the substrate, consequently increasing the adatom density while facilitating longer migration distances. Thus, a substantial increase in the number of adatoms joining the flake can occur. However, under these circumstances, the flake may not have enough time to reshape into a compact triangle. Instead, it may result in defective forms (Figure 2c III and IV) or even branched structures (Figure 2c - V and VI). Equivalent experimental flakes can be found in **Figure S2**. In Figure 2d, we delve into the role of activation energy for on-substrate migration, a parameter that governs the influx of adatoms reaching the crystal edge and, subsequently, their potential incorporation into the crystal. Lowering activation energies for migration facilitates migrations, thereby augmenting the number of atoms available for integrating into the flake. On the other hand, diminished migration rates translate to fewer adatoms participating in flake formation, resulting in slower growth speed and compact triangles.

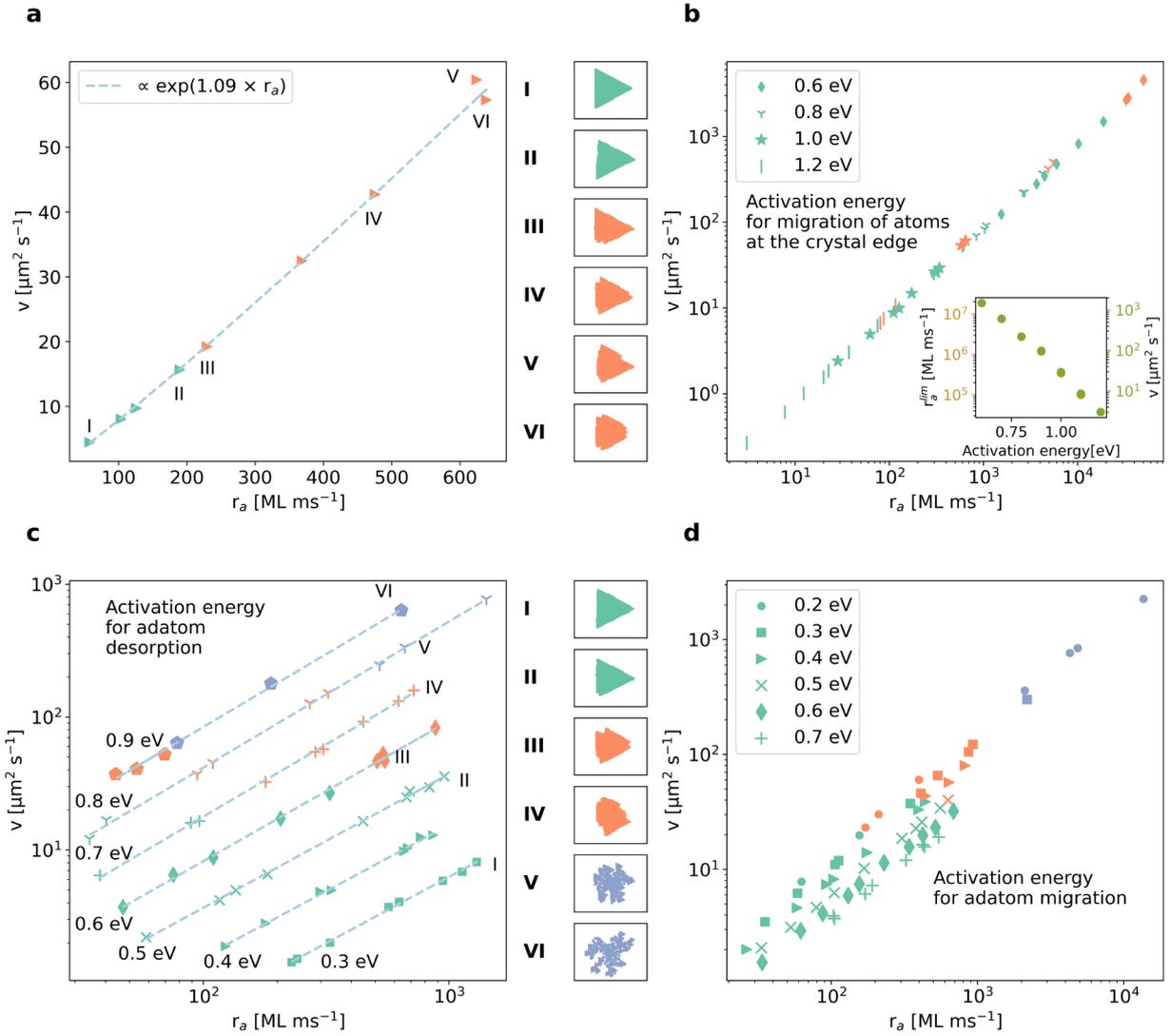

**Figure 2: Single domain with varied growth conditions.** Cyan symbols represent compact triangles, orange symbols denote defective triangles and blue symbols signify branched flakes. Symbol size corresponds to the magnitude of the activation energy, facilitating visual interpretation. A) Depicts the growth speed ($v$) for various adsorption rates ($r_a$). Simulated flakes are shown in the right panel, labelled in accordance with the simulations in the graph. b) Depicts the growth speed ($v$) at different adsorption rates ($r_a$) while employing distinct activation energies for atom migrations along the crystal edge. The inset shows the adsorption rate limit to achieve a compact triangle for various activation energies on the left axis and the corresponding growth speed on the right axis. c) Depicts the growth speed ($v$) at different adsorption rates ($r_a$), considering different activation energies for adatom desorption. Simulated

flakes are shown in the right panel, labeled in accordance with the simulations in the graph. d) Depicts the growth speed ($v$) at different adsorption rates ($r_a$) while employing different activation energies for on-substrate migration.

Engineering the substrate roughness provides us with an effective way to control the flake morphology (see the Supporting Information, section 1, for the experimental details). The substrate surface was patterned into regions with varying roughness. We focus on the border flakes which grow into adjacent regions, i.e., the pristine and etched regions. The area of the border flake on the etched region varies with the roughness of the underlying substrate, and the ratio of the area on the etched region ($S_e$) over the total area ($S_t$), i.e., $r = S_e/S_t$ is shown in **Figure 3a**. To replicate the experimental scenario, we formulate dual-domain simulations featuring a triangular seed located at the border. Specifically, our theoretical analysis focuses on three distinct configurations: symmetric case (refer to the inset in Figure 3b), those oriented towards the shaded region (inset shaded-oriented in Figure 3c) and those oriented towards the unshaded region (inset unshaded-oriented in Figure 3c). It is assumed that increased surface roughness amplifies the adsorption rate ($r_a$), and the shaded and unshaded regions in the insets of Figure 3b and 3c exhibit different adsorption rates. In the symmetric scenario depicted in Figure 3b, we observe that the area ratio between the shaded region and the total flake area is approximately 0.24 when the adsorption ratio is 0.33. The area ratio subsequently increases to 0.72 as the adsorption ratio reaches 1.66, while the growth still results in a compact triangular shape (refer to Figure 3b, I to III). As the adsorption ratio continues to rise, the imbalance between these areas intensifies, resulting in defective flakes (see Figure 3b, IV to VI). The simulation enables us to uncover the growth environment in the experiment by comparing the morphologies of the experimental and simulated flakes. For example, the morphology of the experimental flake in the inset of Figure 3a suggests that the adsorption ratio of the two substrate regions used in the experiment is between 0.66 and 1.66.

Unlike the symmetric flakes, the growth of the flakes in the asymmetric scenario (Figure 3c) exhibits inherent imbalance due to the different growth speeds ($v$) along inequivalent crystal direction. Specifically, flakes oriented towards the unshaded region generally exhibit a smaller area ratio between the unshaded region area ($S_{un}$) and total area, $r = S_{un}/S_t$, compared with those oriented towards the shaded region. This imbalance persists across the range of simulated adsorption rate ratios, although it tends to diminish for the extreme values of adsorption ratios where the flake tends to grow predominantly within one of the regions (see Figure 3c and the accompanying flake morphologies in the right panels). The flake oriented toward the unshaded region maintains a compact shape for most of the adsorption rate ratio range. This compactness is attributed to the continuous and sufficient flux of adatoms to the vertex of the flake from the shaded region. Conversely, for flakes oriented towards the shaded region, the large supply of adatoms from the unshaded region enlarges the triangle base, and the shortage of adatoms at the left vertex eventually results in the distortion and loss of the triangular shape. We observed similar growth behaviours from the experiments (see Figure S3, S4 and S5 in Supporting Information), and our simulation suggests that the adsorption ratio in the experiment is between 1 and 1.66, as over 3.33 we find irregular flakes.
.

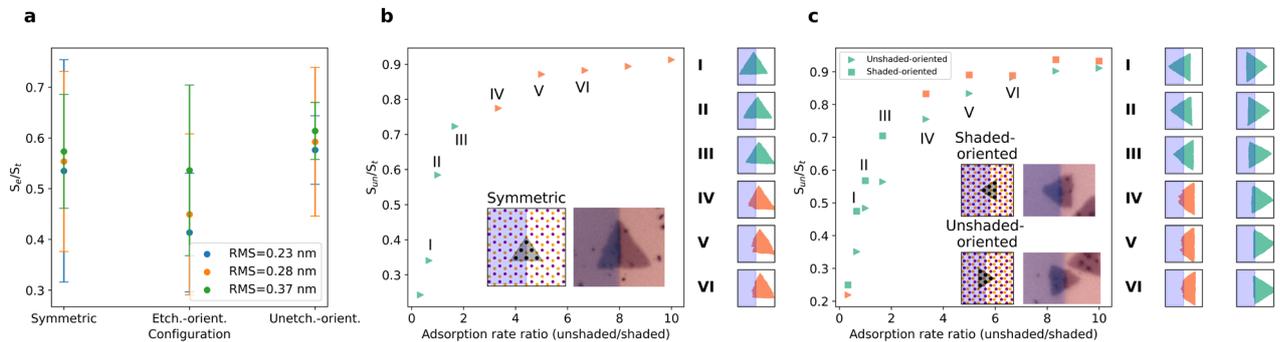

**Figure 3: Dual domain with varying adsorption rates.** Cyan symbols represent compact triangles, while orange symbols denote defective triangles. a) Statistical analysis of experimental flakes situated at the interface of two distinct regions, one of them with RMS

roughness of 0.23, 0.28 and 0.37 nm, while the other remained unaltered. These flakes exhibit three distinct configurations, as depicted in the insets in b) and c). In the experimental context, orange denotes the etched region, while purple represents the unetched region. b) and c) unshaded region ($S_{un}$) over flake area ($S_t$), $r = S_{un}/S_t$, as a function of the adsorption ratio between regions. The simulations are conducted for flakes symmetrically positioned between regions (b) and those asymmetrically oriented towards either the shaded or unshaded region (c).

In **Figure 4**, we vary the activation energies for adatom desorption of the adjacent substrate regions and investigate its effect on the morphology of the border flakes. Specifically, the activation energy of desorption of the shaded region is fixed at 0.6 eV, while the activation energy of the unshaded region is varied from 0.2 to 0.9 eV. We consider the same configurations: flakes oriented towards the unshaded region (Figure 4a and b), flakes oriented towards the shaded region (Figure 4c and d) and flakes symmetrically positioned between these regions (Figure 4e and f). A selection of three flake morphologies corresponding to the labeled points (I, II and III) for each configuration are shown in the right panels. As depicted in Figure 4a and b (unshaded-oriented) and Figure 4e and f (symmetric case), most of the compact triangles are obtained for activation energies for adatom desorption of 0.5, 0.6, and 0.7 eV. In the case of Figure 4c and d (shaded-oriented), we also observe compact triangles for adatom desorption energies below 0.5 eV. This can be attributed to a consistent and sufficient adatom flux toward the flake's vertex from the shaded region, as the adatom desorption energy is higher. This scenario is similar to the one explained in Figure 3c and triangles closely resemble those observed in Figure 3c IV or V (see panel I in Figure 4, middle row). Defective flakes sharing the orientation shown in the inset in Figure 4a typically evolve into trapezoidal shapes, resembling triangles with one vertex missing in the unshaded region (see panel I in the top row). For the symmetric case, illustrated in the inset in Figure 4e, defective triangles tend to adopt a half-triangle shape (see Figure 3b VI). The simulations reveal that even slight modifications in the activation energy for adatom desorption cause profound effects.

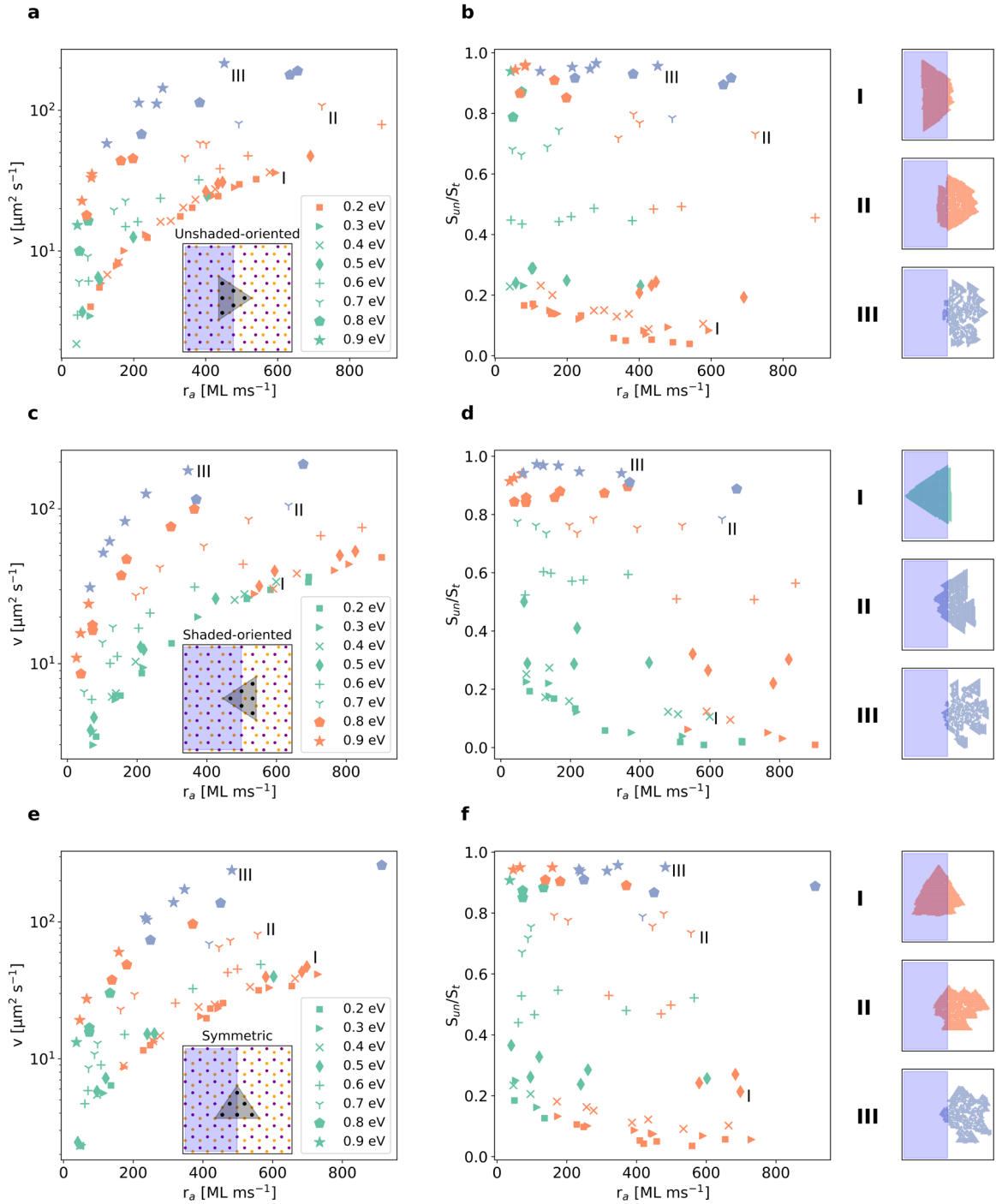

**Figure 4: Dual domain with varying activation energy for desorption.** Cyan symbols represent compact triangles, orange symbols denote defective triangles and blue symbols signify branched flakes. Symbol size corresponds to the magnitude of the activation energy, facilitating visual interpretation. The activation energy for desorption is varied in the unshaded

region and fixed in the shaded. a), c) and e) Depict the growth speed ($v$) at different adsorption rates ($r_a$), considering different activation energies for adatom desorption. b), d) and f) Show the proportion of the unshaded region over the total flake area ($r = S_{un}/S_t$) as a function of the adsorption rates ($r_a$) for different activation energies for desorption. Simulations are conducted for flakes oriented towards the unshaded region (a and b), flakes oriented towards the shaded region (c and d) and flakes symmetrically positioned between regions (e and f).

The impact is twofold, significantly increasing the growth speed ($v$), sometimes by over one order of magnitude, and promoting preferential growth of the flake within one of the regions. Notably, a substantial difference in desorption rates between the regions leads to an adatom density imbalance on either side of the flake. Consequently, distinct growth behaviours manifest within the same flake, an effect that can be mitigated through rapid edge migration. Otherwise, slow edge migrations result in the formation of defective or branched flakes.

**Figure 5** illustrates the impact of varied activation energies for on-substrate adatom migration on flakes growing at the border of two distinct regions. As previously explained in Figure 4, these regions exhibit dissimilar activation energies for adatom migration. The shaded region maintains a constant activation energy of 0.4 eV, while the unshaded region's activation energy is varied. The three configurations are the same as used in the previous figure: flakes oriented towards the unshaded region (Figure 5a and b), flakes oriented towards the shaded region (Figure 5c and d) and flakes symmetrically positioned between these regions (Figure 5e and f). A selection of three flake morphologies corresponding to the labelled points (I, II and III) in Figure 5 can be found in the right panels of each row for each configuration. In contrast to the case involving activation energy for adatom desorption (Figure 4), alterations in the activation energy for on-substrate adatom migration exert a comparatively lesser influence. Lowering energy barriers for migration increases the growth speed ($v$). However, the effect on the area ratio between the unshaded region and the total flake area ($r = S_{un}/S_t$) is less pronounced. Notably, lower activation energies, specifically 0.2 and 0.3 eV, permit adatoms to migrate over longer distances before desorption. Consequently, they can transition more frequently from the unshaded to the shaded region. Additionally, it is observed that for adsorption rates below 100 ML ms$^{-1}$, most triangles retain their compact form across the entire spectrum of investigated activation energies. Unlike scenarios characterized by reduced desorption rates, an increase in migration rate implies a smaller increment in the number of adatoms joining the crystal. Hence, the crystal is afforded additional time for reshaping and maintaining a compact form, leading to a relatively minor increase in growth speed (see Figure 5a, c and e). In Figure 5b, 5d and 5f, we can observe the influence of the activation energy for on-substrate adatom migration on promoting preferential growth in one of the regions. Figure 5b corresponds to the configuration with the triangle oriented towards the unshaded region, where the proportion of the unshaded region over the total flake area ($r = S_{un}/S_t$) starts with a value lower than 0.5, as the base of the triangle lies in the shaded region. We observe that for the entire range of activation energies, the values remain under 0.5. Conversely, we find the opposite case in the configuration of the triangle oriented towards the shaded region in Figure 5d. In this instance, most of the area ratios are over 0.5, with a few exceptions where low activation energies permit adatoms to migrate to the shaded region and contribute to that area. For the triangles situated symmetrically between the regions in Figure 5f, we find that most of the cases are randomly distributed around 0.5, with some skewness towards the shaded region (values exceeding 0.5). Consequently, modulating the on-substrate adatom migration rate does not appear to be highly effective in controlling the region where the flake preferably grows, in contrast to the case of activation energy for adatom desorption shown in Figure 4. However, in this scenario, a greater number of triangles maintain a compact form.

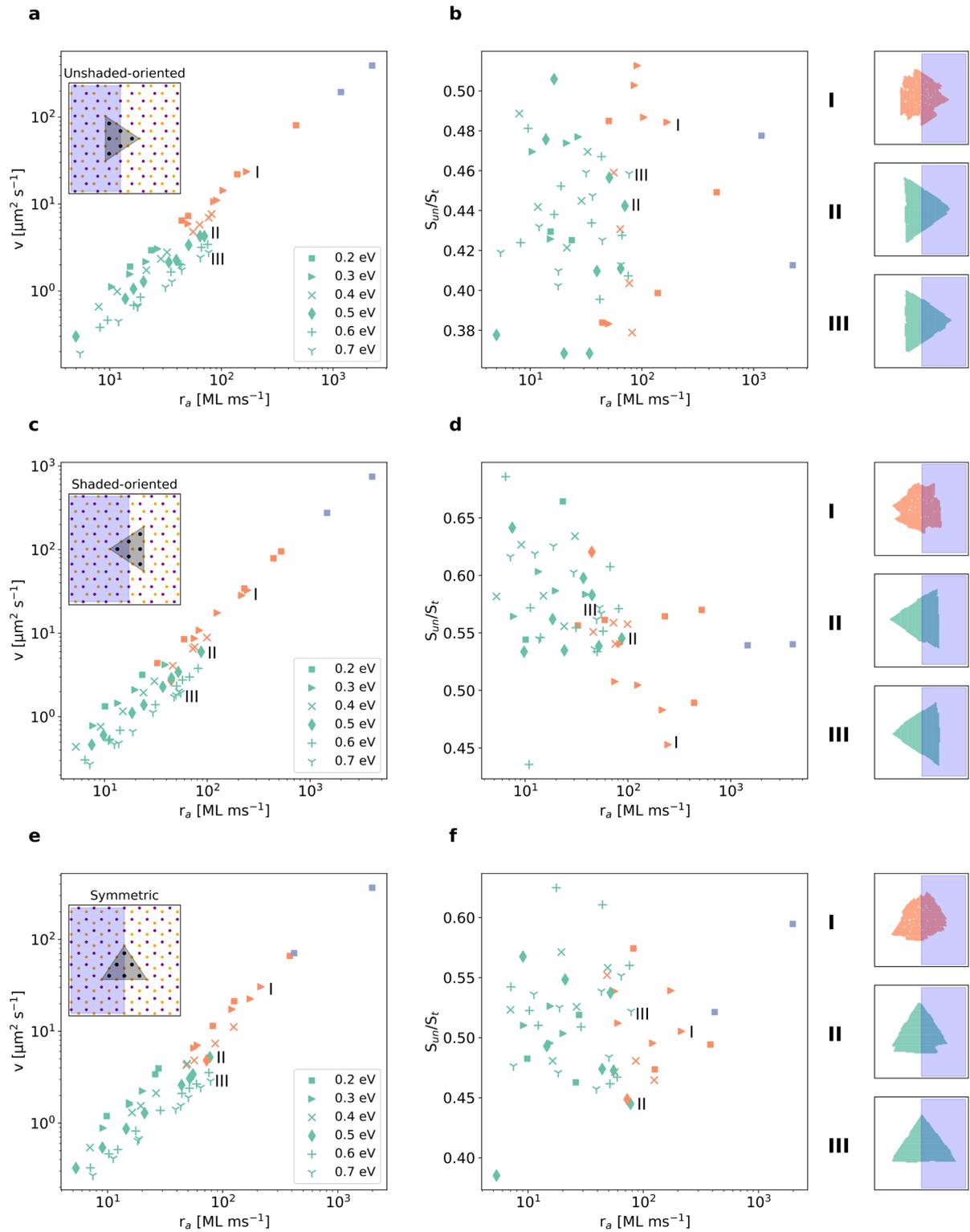

**Figure 5: Double domain with varying activation energy for on-substrate migration.** Cyan symbols represent compact triangles, orange symbols denote defective triangles and blue

symbols signify branched flakes. Symbol size corresponds to the magnitude of the activation energy, facilitating visual interpretation. The activation energy for on-substrate adatom migration is varied in the unshaded region and fixed in the shaded. a), c) and e) Depict the growth speed ($v$) at different adsorption rates ($r_a$), considering different activation energies for on-substrate adatom migration. b), d) and f) Show the proportion of the unshaded region over the total flake area ($r = S_{un}/S_t$) as a function of the adsorption rates ($r_a$) for different activation energies for on-substrate adatom migration. Simulations are conducted for flakes oriented towards the unshaded region (a and b), flakes oriented towards the shaded region (c and d) and flakes symmetrically positioned between regions (e and f).

**Conclusion**

In this study, we employed rejection-free kinetic Monte Carlo physical simulations based on a hexagonal on-lattice diffusion model to delve into the intricate dynamics of Transition Metal Dichalcogenides flake morphology during the growth process. We explore the interplay between various parameters, including adsorption rates and activation energies for on-substrate adatom migration, edge migration, and adatom desorption. Our simulations centered on modeling triangular MoS$_2$ flakes on various substrates, as well as flakes growing at the interface of two different substrates. The simulations show the significant impact of adsorption rates on growth kinetics. While elevated adsorption rates expedite growth, they simultaneously engendered the formation of defective flakes. We find that using adsorption ratios between regions below 1.66 presented a viable strategy for steering preferential growth in a specific region while maintaining a compact shape. Crucially, the balance between the incorporation of adatoms into the crystal and edge migration emerged as a pivotal determinant of flake morphology, dictating whether the final structure assumed a compact, defective, or branched form. Furthermore, we examined that small variations in the activation energy for adatom desorption have a pronounced impact on growth speed and promotion of preferential growth in one of the regions even for small differences between regions. Notably, small differences between desorption rates are needed to achieve compact triangular flakes. We also explore the influence of activation energies regulating on-substrate adatom migration concluding that have a relatively modest effect on the number of adatoms joining the crystal compared with reduced desorption rates. Consequently, this affords the crystal an extended time for structural adaptation, resulting in a relatively modest increase in growth speed. In summary, our study provides valuable insights into the complex flake morphology dynamics. The study sheds light on the potentials and limitations of the different processes governing the growth of $MoS_2$ flakes. Importantly, the knowledge extracted from this work can be also applied to other TMDs. Furthermore, the magnitudes used are tied to substrate properties and the quantity of reactants employed in experimental setups. Consequently, this knowledge helps to make decisions in designing and optimizing TMD synthesis processes.

**Supporting Information**

Supporting Information is available from the Wiley Online Library or from the author.

**Author Contributions**

S. A. developed the simulator and conducted the simulation. S. A., I. S., L. W. and H. Z. analyzed the data and wrote the manuscript. H.Z. supervised the project. L. W. and I. S. growth the samples and processed the experimental data. All authors have given approval to the final version of the manuscript.

**Acknowledgements**

The authors gratefully acknowledge financial support by the Science Foundation Ireland under 20/FFP-P/8727, Irish Research Council under EPSPG/2020/81 and the support of the Irish

National e-Infrastructure (e-INIS). All calculations were performed on the Kelvin cluster maintained by the Trinity Centre for High Performance Computing. This cluster was funded through grants from the Higher Education Authority, through its PRTLI program.

Received: ((will be filled in by the editorial staff))
Revised: ((will be filled in by the editorial staff))
Published online: ((will be filled in by the editorial staff))

**References**
[1]	J. A. Wilson, A. D. Yoffe, *Adv. Phys.* **1969**, 18, 193.
[2]	K. F. Mak, C. Lee, J. Hone, J. Shan, T. F. Heinz, *Phys. Rev. Lett.* **2010**, 105, 4, 136805.
[3]	B. Radisavljevic, A. Radenovic, J. Brivio, V. Giacometti, A. Kis, *Nat. Nanotechnol.* **2011**, 6, 147.
[4]	A. Splendiani, L. Sun, Y. B. Zhang, T. S. Li, J. Kim, C. Y. Chim, G. Galli, F. Wang, *Nano Lett.* **2010**, 10, 1271.
[5]	V. K. Sangwan, D. Jariwala, I. S. Kim, K. S. Chen, T. J. Marks, L. J. Lauhon, M. C. Hersam, *Nat. Nanotechnol.* **2015**, 10, 403.
[6]	V. K. Sangwan, H. S. Lee, H. Bergeron, I. Balla, M. E. Beck, K. S. Chen, M. C. Hersam, *Nature* **2018**, 554, 500.
[7]	J. Jadwiszczak, D. Keane, P. Maguire, C. P. Cullen, Y. B. Zhou, H. D. Song, C. Downing, D. Fox, N. McEvoy, R. Zhu, J. Xu, G. S. Duesberg, Z. M. Liao, J. J. Boland, H. Z. Zhang, *ACS Nano* **2019**, 13, 14262.
[8]	D. Li, B. Wu, X. J. Zhu, J. T. Wang, B. Ryu, W. D. Lu, W. Lu, X. G. Liang, *ACS Nano* **2018**, 12, 9240.
[9]	S. Kim, A. Konar, W. S. Hwang, J. H. Lee, J. Lee, J. Yang, C. Jung, H. Kim, J. B. Yoo, J. Y. Choi, Y. W. Jin, S. Y. Lee, D. Jena, W. Choi, K. Kim, *Nat. Commun.* **2012**, 3, 7, 1011.
[10]	S. Bertolazzi, P. Bondavalli, S. Roche, T. San, S. Y. Choi, L. Colombo, F. Bonaccorso, P. Samori, *Adv. Mater.* **2019**, 31, 35, 1806663.
[11]	W. Z. Wu, L. Wang, Y. L. Li, F. Zhang, L. Lin, S. M. Niu, D. Chenet, X. Zhang, Y. F. Hao, T. F. Heinz, J. Hone, Z. L. Wang, *Nature* **2014**, 514, 470.
[12]	Y. F. Nie, C. P. Liang, P. R. Cha, L. G. Colombo, R. M. Wallace, K. Cho, *Sci Rep* **2017**, 7, 13, 2977.
[13]	H. Fang, S. Chuang, T. C. Chang, K. Takei, T. Takahashi, A. Javey, *Nano Lett.* **2012**, 12, 3788.
[14]	H. J. Chuang, X. B. Tan, N. J. Ghimire, M. M. Perera, B. Chamlagain, M. M. C. Cheng, J. Q. Yan, D. Mandrus, D. Tomanek, Z. X. Zhou, *Nano Lett.* **2014**, 14, 3594.
[15]	H. Tian, Z. Tan, C. Wu, X. M. Wang, M. A. Mohammad, D. Xie, Y. Yang, J. Wang, L. J. Li, J. Xu, T. L. Ren, *Sci Rep* **2014**, 4, 9, 5951.
[16]	R. Y. Yue, A. T. Barton, H. Zhu, A. Azcatl, L. F. Pena, J. Wang, X. Peng, N. Lu, L. X. Cheng, R. Addou, S. McDonnell, L. Colombo, J. W. P. Hsu, J. Kim, M. J. Kim, R. M. Wallace, C. L. Hinkle, *ACS Nano* **2015**, 9, 474.
[17]	H. C. Diaz, R. Chaghi, Y. J. Ma, M. Batzill, *2D Mater.* **2015**, 2, 10, 044010.

[18]	S. M. Eichfeld, L. Hossain, Y. C. Lin, A. F. Piasecki, B. Kupp, A. G. Birdwell, R. A. Burke, N. Lu, X. Peng, J. Li, A. Azcatl, S. McDonnell, R. M. Wallace, M. J. Kim, T. S. Mayer, J. M. Redwing, J. A. Robinson, *ACS Nano* **2015**, 9, 2080.
[19]	J. K. Huang, J. Pu, C. L. Hsu, M. H. Chiu, Z. Y. Juang, Y. H. Chang, W. H. Chang, Y. Iwasa, T. Takenobu, L. J. Li, *ACS Nano* **2014**, 8, 923.
[20]	G. Deokar, J. Avila, I. Razado-Colambo, J. L. Codron, C. Boyaval, E. Galopin, M. C. Asensio, D. Vignaud, *Carbon* **2015**, 89, 82.
[21]	T. S. Millard, A. Genco, E. M. Alexeev, S. Randerson, S. Ahn, A. R. Jang, H. S. Shin, A. I. Tartakovskii, *npj 2D Mater. Appl.* **2020**, 4, 12.
[22]	S. Najmaei, M. Amani, M. L. Chin, Z. Liu, A. G. Birdwell, T. P. O'Regan, P. M. Ajayan, M. Dubey, J. Lou, *ACS Nano* **2014**, 8, 7930.
[23]	T. H. Ly, D. J. Perello, J. Zhao, Q. Deng, H. Kim, G. H. Han, S. H. Chae, H. Y. Jeong, Y. H. Lee, *Nat. Commun.* **2016**, 7, 10426.
[24]	T. Kwak, J. Lee, B. So, U. Choi, O. Nam, *J. Cryst. Growth* **2019**, 510, 50.
[25]	W. Hong, C. Park, G. W. Shim, S. Y. Yang, S. Y. Choi, *ACS Appl. Mater. Interfaces* **2021**, 13, 50497.
[26]	A. M. van der Zande, P. Y. Huang, D. A. Chenet, T. C. Berkelbach, Y. M. You, G. H. Lee, T. F. Heinz, D. R. Reichman, D. A. Muller, J. C. Hone, *Nat. Mater.* **2013**, 12, 554.
[27]	Y. F. Nie, C. P. Liang, K. H. Zhang, R. Zhao, S. M. Eichfeld, P. R. Cha, L. Colombo, J. A. Robinson, R. M. Wallace, K. Cho, *2D Mater.* **2016**, 3, 10, 025029.
[28]	L. X. Wu, W. H. Yang, G. F. Wang, *npj 2D Mater. Appl.* **2019**, 3, 7, 6.
[29]	G. Ghimire, R. K. Ulaganathan, A. Tempez, O. Ilchenko, R. R. Unocic, J. Heske, D. I. Miakota, C. Xiang, M. Chaigneau, T. Booth, P. Boggild, K. S. Thygesen, D. B. Geohegan, S. Canulescu, *Adv. Mater.* **2023**, 35, 11.
[30]	S. Chen, J. F. Gao, B. M. Srinivasan, Y. W. Zhang, *Acta Phys.-Chim. Sin.* **2019**, 35, 1119.
[31]	S. S. Wang, Y. M. Rong, Y. Fan, M. Pacios, H. Bhaskaran, K. He, J. H. Warner, *Chem. Mat.* **2014**, 26, 6371.
[32]	L. X. Wu, W. H. Yang, G. F. Wang, *npj 2D Mater. Appl.* **2019**, 3, 7, 6.
[33]	Y. Z. Ji, K. Momeni, L. Q. Chen, *2D Mater.* **2021**, 8, 14, 035033.
[34]	D. Cao, T. Shen, P. Liang, X. S. Chen, H. B. Shu, *J. Phys. Chem. C* **2015**, 119, 4294.
[35]	Y. Zhang, Q. Q. Ji, G. F. Han, J. Ju, J. P. Shi, D. L. Ma, J. Y. Sun, Y. S. Zhang, M. J. Li, X. Y. Lang, Y. F. Zhang, Z. F. Liu, *ACS Nano* **2014**, 8, 8617.
[36]	N. Briggs, S. Subramanian, Z. Lin, X. F. Li, X. T. Zhang, K. H. Zhang, K. Xiao, D. Geohegan, R. Wallace, L. Q. Chen, M. Terrones, A. Ebrahimi, S. Das, J. Redwing, C. Hinkle, K. Momeni, A. van Duin, V. Crespi, S. Kar, J. A. Robinson, *2D Mater.* **2019**, 6, 23, 022001.
[37]	K. Momeni, Y. Z. Ji, K. H. Zhang, J. A. Robinson, L. Q. Chen, *npj 2D Mater. Appl.* **2018**, 2, 7, 27.
[38]	K. Momeni, Y. Z. Ji, Y. X. Wang, S. Paul, S. Neshani, D. E. Yilmaz, Y. K. Shin, D. F. Zhang, J. W. Jiang, H. S. Park, S. Sinnott, A. van Duin, V. Crespi, L. Q. Chen, *npj Comput. Mater.* **2020**, 6, 18, 22.
[39]	Z. N. Qi, P. H. Cao, H. S. Park, *J. Appl. Phys.* **2013**, 114, 4, 163508.
[40]	H. Qiu, T. Xu, Z. L. Wang, W. Ren, H. Y. Nan, Z. H. Ni, Q. Chen, S. J. Yuan, F. Miao, F. Q. Song, G. Long, Y. Shi, L. T. Sun, J. L. Wang, X. R. Wang, *Nat. Commun.* **2013**, 4, 6, 2642.
[41]	S. Y. Zhu, Q. Wang, *AIP Adv.* **2015**, 5, 8, 107105.
[42]	J. V. Lauritsen, M. V. Bollinger, E. Laegsgaard, K. W. Jacobsen, J. K. Norskov, B. S. Clausen, H. Topsoe, F. Besenbacher, *J. Catal.* **2004**, 221, 510.
[43]	S. Paul, R. Torsi, J. A. Robinson, K. Momeni, *ACS Appl. Mater. Interfaces* **2022**, 14, 18835.

[44] K. Momeni, Y. Z. Ji, N. Nayir, N. Sakib, H. Y. Zhu, S. Paul, T. H. Choudhury, S. Neshani, A. C. T. van Duin, J. M. Redwing, L. Q. Chen, *npj Comput. Mater.* **2022**, 8, 8, 240.
[45] K. Momeni, Y. Z. Ji, L. Q. Chen, *J. Mater. Res.* **2022**, 37, 114.
[46] A. Chatterjee, D. G. Vlachos, *J. Comput-Aided Mater. Des.* **2007**, 14, 253.
[47] C. C. Battaile, *Comput. Meth. Appl. Mech. Eng.* **2008**, 197, 3386.
[48] A. F. Voter, presented at *Conference of the NATO-Advanced-Study-Institute on Radiation Effects in Solids*, Erice, ITALY, Jul 17-29, **2004**.
[49] S. Ovesson, A. Bogicevic, B. I. Lundqvist, *Phys. Rev. Lett.* **1999**, 83, 2608.
[50] R. Whitesides, M. Frenklach, *J. Phys. Chem. A* **2010**, 114, 689.
[51] S. Aldana, J. Jadwiszczak, H. Zhang, *Nanoscale* **2023**.
[52] S. Aldana, H. Z. Zhang, *ACS Omega* **2023**, 8, 27543.
[53] H. Shu, X. Chen, X. Tao, F. Ding, *ACS Nano* **2012**, 6, 3243.
[54] Y. Gao, Y.-L. Hong, L.-C. Yin, Z. Wu, Z. Yang, M.-L. Chen, Z. Liu, T. Ma, D.-M. Sun, Z. Ni, X.-L. Ma, H.-M. Cheng, W. Ren, *Adv. Mater.* **2017**, 29, 1700990.